\documentclass[12pt]{iopart}

\usepackage{color}
\usepackage{iopams}
\usepackage{mathrsfs}
\usepackage{latexsym}
\usepackage{subfigure, graphicx}
\usepackage{caption}
\usepackage{float}
\usepackage{ulem}

\begin{document}

\def\prd{{\it Phys. Rev.} D~}
\def\PRL{{\it Phys.Rev.} Lett~}
\def\apjl{{\it Astrophys. J.} Lett~}
\def\apj{{\it Astrophys. J.}}
\def\Msun{M_\odot}
\def\PRD{{\it Phys. Rev.} D~}
\def\CQG{{\it Class. Quantum Grav.}}
\def\aaps{{\it A\&AS~}}
\def\pasj{{\it PASJ~}}
\def\mnras{{\it MNRAS~}} 
\def\aapr{{\it A\&ARv~}}
\def\aap{{\it A\&A~}}
\def\araa{{\it A\&RAA~}}
\def\na{{\it New Astronomy~}}
\def\ptp{{\it Progress of Theoretical Physics~}}

\newcommand{\rh}[1]{\textcolor{blue}{RH: #1}}
\newcommand{\dsj}[1]{\textcolor{green}{DJ: #1}}
\newcommand{\el}[1]{\textcolor{red}{EL: #1}}
		
	\title[POWER: Open source software to post-process numerical relativity simulations]{\texttt{P}ython \texttt{O}pen Source \texttt{W}aveform \texttt{E}xtracto\texttt{R} (\texttt{POWER}): An open source, \texttt{Python} package to monitor and post-process numerical relativity simulations}

	\author{Daniel Johnson$^{1, 2, 3, 4}$, E.~A. Huerta$^1$, and Roland Haas$^1$}
	\address{$^1$ NCSA, University of Illinois at Urbana-Champaign, Urbana, Illinois, 61801}
	\address{$^2$ Department of Physics, University of Illinois at Urbana-Champaign, Urbana, Illinois, 61801}
	\address{$^3$ Department of Computer Science, University of Illinois at Urbana-Champaign, Urbana, Illinois, 61801}
	\address{$^4$ Students Pushing Innovation (SPIN) undergraduate intern at NCSA, University of Illinois at Urbana-Champaign, Urbana, Illinois 61801, USA}

	\vspace{10pt}
	\begin{indented}
		\item[\today]
	\end{indented}
	
	\begin{abstract}
		Numerical simulations of Einstein's field equations provide unique insights into the physics of compact objects moving at relativistic speeds, and which are driven by strong gravitational interactions. Numerical relativity has played a key role to firmly establish gravitational wave astrophysics as a new field of research, and it is now paving the way to establish whether gravitational wave radiation emitted from compact binary mergers is accompanied by electromagnetic and astro-particle counterparts. As numerical relativity continues to blend in with routine gravitational wave data analyses to validate the discovery of gravitational wave events, it is essential to develop open source tools to streamline these studies. Motivated by our own experience as users and developers of the open source, community software, the \texttt{Einstein Toolkit}, we present an open source, \texttt{Python} package that is ideally suited to monitor and post-process the data products of numerical relativity simulations, and compute the gravitational wave strain at future null infinity in high performance environments. We showcase the application of this new package to post-process a large numerical relativity catalog and extract higher-order waveform modes from numerical relativity simulations of eccentric binary black hole mergers and neutron star mergers. This new software fills a critical void in the arsenal of tools provided by the \texttt{Einstein Toolkit} Consortium to the numerical relativity community.
	\end{abstract}
	
	\pacs{07.05.Mh, 07.05.Kf, 04.80.Nn, 95.55.Ym}
	%
	\vspace{2pc}
	\noindent{\it Keywords}: LIGO, Gravitational Waves, Numerical Relativity

	%
	%
	%
	%

	\section{Introduction}
	\label{intro}
	Einstein's theory of general relativity is one of the greatest accomplishments in theoretical physics~\cite{gr}. The study of this theory has led to the creation of several research programs in mathematics, experimental and theoretical physics, astronomy, and computer science. Analytical studies of this theory by mathematicians and theoretical astrophysicists over the last century have shed light into the physics of new astrophysical objects, and processes that are dominated by strong gravitational interactions. The advent of fully numerical solutions of Einstein's field equations imparted a major boost to the study of gravitational wave (GW) sources~\cite{preto,naka:1987,shiba:1995,baum:1999,baker:2006,camp:2006}, i.e., mergers of compact objects, such as binary black holes (BBHs), binary neutron stars (BNSs), supernovae explosions, etc~\cite{SathyaLRR:2009}.
	
	Numerical relativity (NR) paved the way for the development of tools that were critical for the detection of GWs with the twin Advanced Laser Interferometer Gravitational Wave Observatory (aLIGO) detectors~\cite{DI:2016,secondBBH:2016,thirddetection,2017arXiv170909660T}. In addition to providing NR waveforms for the calibration of waveform models that were used extensively to detect and characterize GW transients in aLIGO data~\cite{Tara:2014,khan:2016PhRvD}, NR simulations are also directly used in aLIGO data analyses to validate the astrophysical origin of GW transients~\cite{2016arXiv161107531T,NRI:2016}. 
	
	NR will continue to shed light on the physics of compact binary mergers that are putative sources for the generation of the most energetic explosions in the Universe~\cite{Lehner:2014a,scenarioligo:2016LRR}. To accomplish these major scientific milestones, numerical relativists are developing software to simulate mergers of NSs and NSBH systems in numerical settings that reproduce, with ever increasing realism, actual astrophysical scenarios~\cite{2017JCoPh.335...84K,2016PhRvD..93l4062H,2015CQGra..32q5009E,2014CQGra..31a5005M,ETL:2012CQGra}. While some of this code remains closed source, there have been community efforts to develop, maintain, and exploit open source NR software, such as the Wolfram Language based package \texttt{SimulationTools}~\cite{simulationtools-web}. In this article, we further the development of open source software for the study of GW sources, which will be included in upcoming releases of the open source, community software, the \texttt{Einstein Toolkit}~\cite{ETL:2012CQGra,etweb}.
	
	The software we introduce in this article addresses a critical void in the arsenal of open source tools to monitor and post-process the data products of large scale NR campaigns in high performance computing (HPC) environments. 
In our experience running and post-processing NR simulations, we have found two paradigms: (i) members of NR groups that develop closed source NR software have a comprehensive toolkit for end-to-end analyses, from the generation of initial data to the extraction of the waveform strain at future null infinity. These tools are not available to the NR community at large; (ii) users of open source software, such as the \texttt{Einstein Toolkit}, develop in-house software for initial data generation, and post-processing of NR simulations. These software packages have not been integrated in code releases of the \texttt{Einstein Toolkit}. One of the existing tools available to post-process NR simulations is \texttt{SimulationTools}\cite{simulationtools-web}, an open source package written in \texttt{Wolfram Language}. We have explored the use of this package and have found that 

\begin{itemize}
\item Purchasing \texttt{Mathematica} licenses can create a barrier to entry users
\item Even though HPC providers allow the use of commercial software, e.g., \texttt{Wolfram Language}, as long as users bring their own licenses, using \texttt{SimulationTools} demands a major overhaul of existing libraries installed cluster wide on HPC facilities such as XSEDE and the Blue Waters supercomputer. This is a rather unfeasible prospect for a narrow application of code based on commercial software
\item If one wanted to use \texttt{SimulationTools} to post-process the data products of a NR campaign run using Blue Waters, XSEDE or any other supercomputing facility that is not compatible with \texttt{Wolfram Language}, one is required to transfer simulation data to an environment compatible with \texttt{Wolfram Language}, which then leads to unnecessary duplication of large datasets.
\end{itemize}

\noindent In this note, we address these issues by introducing a new open source \texttt{Python} package that facilitates and streamlines the monitoring and post-processing of large scale NR campaigns on any platform, using available software modules on HPC environments. 
	
	This paper is organized as follows: in Section~\ref{post}, we introduce \texttt{POWER} (\texttt{P}ython \texttt{O}pen Source \texttt{W}aveform \texttt{E}xtracto\texttt{R}), describe how to obtain this package, and provide a brief description of its usage and features. In Section~\ref{compare}, we use \texttt{POWER} and \texttt{SimulationTools} to compute the GW strain at future null infinity of a number of \texttt{Einstein Toolkit} NR simulations, that we generated using the Blue Waters supercomputer. Based on these results, we show that \texttt{POWER} provides a ready-to-use solution to monitor NR campaigns and post-process their data products on HPC environments. In Section~\ref{ext} we showcase the application of \texttt{POWER} for three timely research topics in the context of vacuum and matter NR simulations. We describe future directions of work in Section~\ref{end}.

	\section{Post-processing of numerical relativity simulations in high performance environments}
	\label{post}
	
	Among the data products of a typical NR simulation, we obtain a set of \texttt{HDF5}\cite{hdf5-web} files that contain numerical solutions for the Weyl scalar \(\Psi_4 (t)\), which is related to the GW strain \(h(t)\) as follows
	
	\begin{eqnarray}
	\label{h_exp}
	h&=& h_{+}- i h_{\times} = \int_{-\infty}^{t} \mathrm{d} \tau' \int_{-\infty}^{\tau'}\mathrm{d} t' \,\Psi_4 \left(t'\right)\,.
	\end{eqnarray}

	\noindent In the transverse traceless gauge, (\(h_{+},\, h_{\times})\) are the two polarizations of the strain, and represent the observables that can be measured with a GW detector. In order to compute \(h(t)\) at null infinity---where GWs are formally defined---it is necessary to convert \(\Psi_4 (t)\) into \(h(t)\), and then extrapolate the strain to infinite radius. This subtle and delicate numerical procedure is done as follows~\cite{reis:2011CQG}. 
	
	We convert the \(\Psi_4\) simulation data to strain \(h(t)\) at each extraction radii using fixed frequency integration (FFI). The cutoff frequency for FFI is obtained from initial parameters by computing the post-Newtonian orbital frequency for the given separation and angular momentum, and then setting this as the cutoff frequency. 
	
	To extrapolate the waveform strain to infinite radius, we convert \(\Psi_4\) to strain at each extraction radii; interpolate the phase, \(\phi\), and amplitude, \({\cal{A}}\), of the strain at each extraction radii to a fixed retarded time \(t_{\rm ret} = t-r\). Finally, we fit an ansatz \(f(r) = f_{\rm inf} + A/r + B/r^2\) to the phase, and  \(r\times {\cal{A}}\) for the amplitude, respectively. For the phase we fix \(B=0\), while for the amplitude $(f_{\rm inf},\, A,\, B\)) are fitted for. POWER will fail with an error message in the case where the \(\Psi_4\) amplitude goes to zero.
	
	Having described the standard method for waveform extraction, we now describe how to obtain and use \texttt{POWER}. Our software is available at NCSA's public gitlab repository, and can be cloned as follows
	
	\vspace{2mm}
	
	\noindent \small{\texttt{REPO$\equiv$\texttt{https://git.ncsa.illinois.edu/elihu/Gravitational\_Waveform\_Extractor.git}}}
	
	\vspace{2mm}
	
	\noindent \$ \texttt{git clone REPO}
	
	\vspace{2mm}
	
	\noindent The command line arguments required to compute the waveform strain \(h(t)\) at future null infinity ares:
	
\vspace{2mm}

\noindent \$ \texttt{cd Waveform\_Extractor/POWER}\\
	\$ \texttt{python power.py} \texttt{n\_radii} \texttt{sim\_1} \texttt{sim\_2} \dots
	
\vspace{2mm}
	
	\noindent where \texttt{sim\_1} and \texttt{sim\_2} represent the NR simulations from which \(h(t)\) is extracted using multiple CPU cores if available, and \texttt{n\_radii} specifies how many of the innermost extraction radii are used for the extrapolation. If \texttt{n\_radii} is not provided, then all extraction radii are used. The extraction radii are given in the parameter file used to run NR simulations with the \texttt{Einstein Toolkit}. Upon executing the above script, the directory, `\texttt{Extrapolated\_Strain}', is automatically created and the files containing \(h(t)\) computed at future null infinity are  written therein as \texttt{ASCII} time series. The files containing \(h(t)\) at future null infinity are labelled as \texttt{simulation\_name\_radially\_extrapolated\_strain.dat}. The repository containing this software includes a sample simulation J0040\_N40. For this specific case, we can compute \(h(t)\) as follows:

\vspace{2mm}

\noindent \$ \texttt{python power.py} 5 \texttt{./simulations/J0040\_N40}

\vspace{2mm}

\noindent which will compute \(h(t)\) using the five innermost extraction radii. To display the plot of the strain \(h(t)\), we use:

\vspace{2mm}

\noindent \$ \texttt{python plot.py J0040\_N40} 

\vspace{2mm}

\noindent To monitor the status of a given set of simulations, \texttt{POWER} provides the following tools. 

The execution of the script:

\vspace{2mm}

\noindent \$ \texttt{monitor\_waveform.sh} \texttt{sim\_1} \texttt{sim\_2} \dots
	
\vspace{2mm}

\noindent generates two separate plots simultaneously, namely, the waveform strain and the waveform phase. Another diagnostics tool provided by \texttt{POWER} is the computation of the radiated energy and angular momentum, which we compute directly from the waveform strains and their time derivatives. We also enable the user to compare these quantities to ADM mass and angular momentum versus time inside a given gravitational waveform extraction radius.

	Having access to this information while a NR campaign is conducted is critical to determine whether the parameter space and resolutions chosen to study an astrophysical problem are sufficient and adequate, respectively. We have found that extracting this information from NR campaigns in an interactive manner saves times and maximizes the use of computational resources. In the following section, we present a direct comparison between \texttt{POWER} and \texttt{SimulationTools} for a variety of NR simulations that are part of the NCSA Numerical Relativity Catalog of Eccentric Compact Binary Coalescence, which is presented in an accompanying publication~\cite{Huerta:ncsacatalog}.

	\section{Comparison to \texttt{SimulationTools}}
	\label{compare}
	
	In this section we compute \(h(t)\) at future null infinity using \texttt{POWER}, and directly compare it to \texttt{SimulationTools}.  \Tref{sims} lists the simulations taken from our NCSA Catalog of NR simulations~\cite{Huerta:ncsacatalog} for this analysis.

	\begin{table}[H]
		\caption{\label{results}Simulations taken from the NCSA Catalog of Eccentric Compact Binary Coalescence~\cite{Huerta:ncsacatalog}. \((e,\, \ell)\) represent the eccentricity and mean anomaly, respectively.}
		\footnotesize
		\begin{center}
			\begin{tabular}{@{}c c c c c c} 
				\hline 
				Simulation & mass-ratio & $e$ & $\ell$ \\ 
				\hline
				E0001 & 1.0 & 0.069 & 3.038 \\ 
				\hline 
				E0017 & 3.0 & 0.073 & 2.957 \\ 
				\hline 
				E0021 & 3.5 & 0.075 & 2.864 \\ 
				\hline 
				J0068 & 4.5 & 0.199& 2.900\\ 
				\hline 
			\end{tabular} 
		\end{center}
		\label{sims}
	\end{table}
	\normalsize

		 	\begin{figure*}[htp!]
			\centerline{
				\includegraphics[height=0.4\textwidth,  clip]{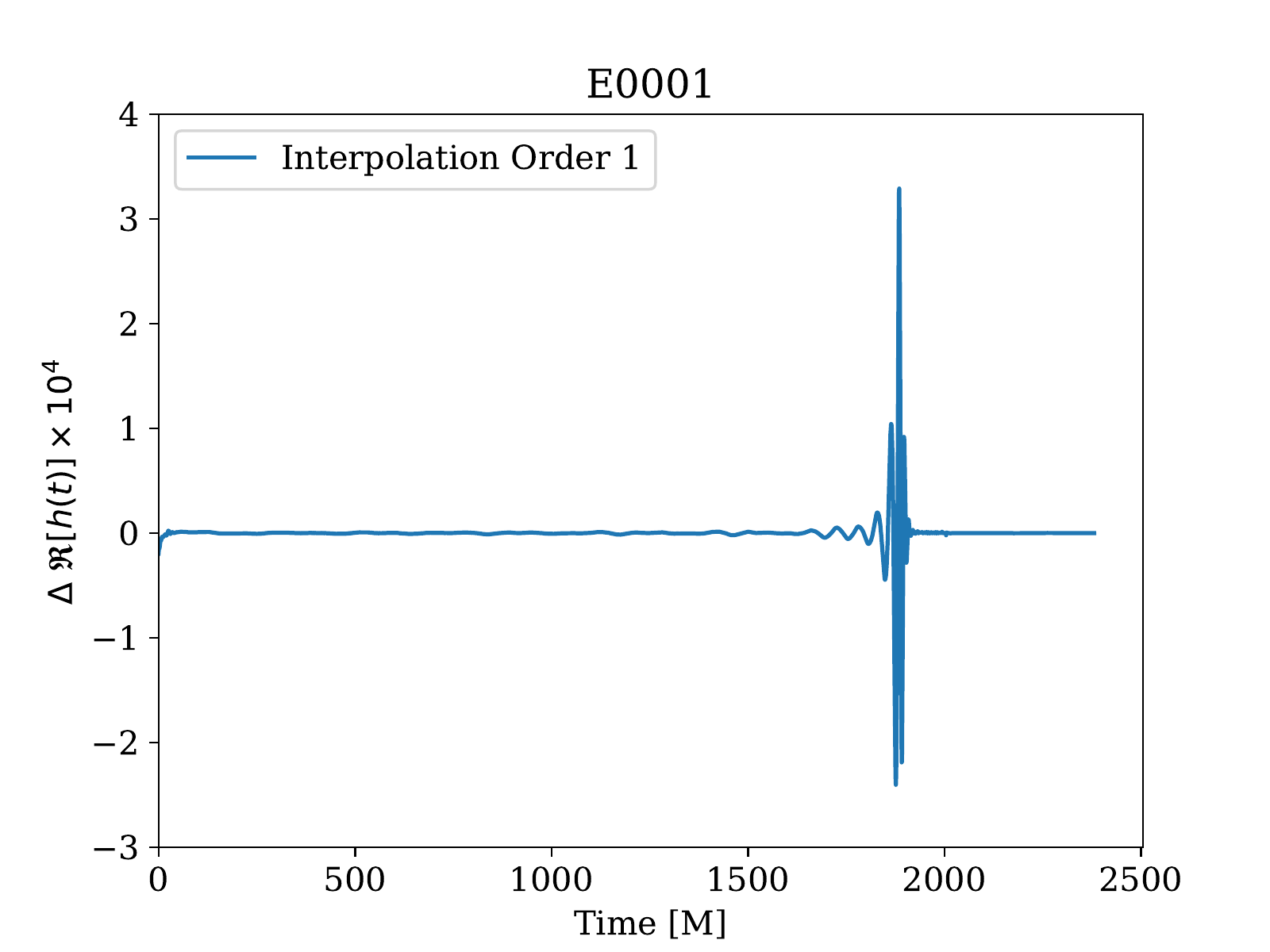}
				\includegraphics[height=0.4\textwidth,  clip]{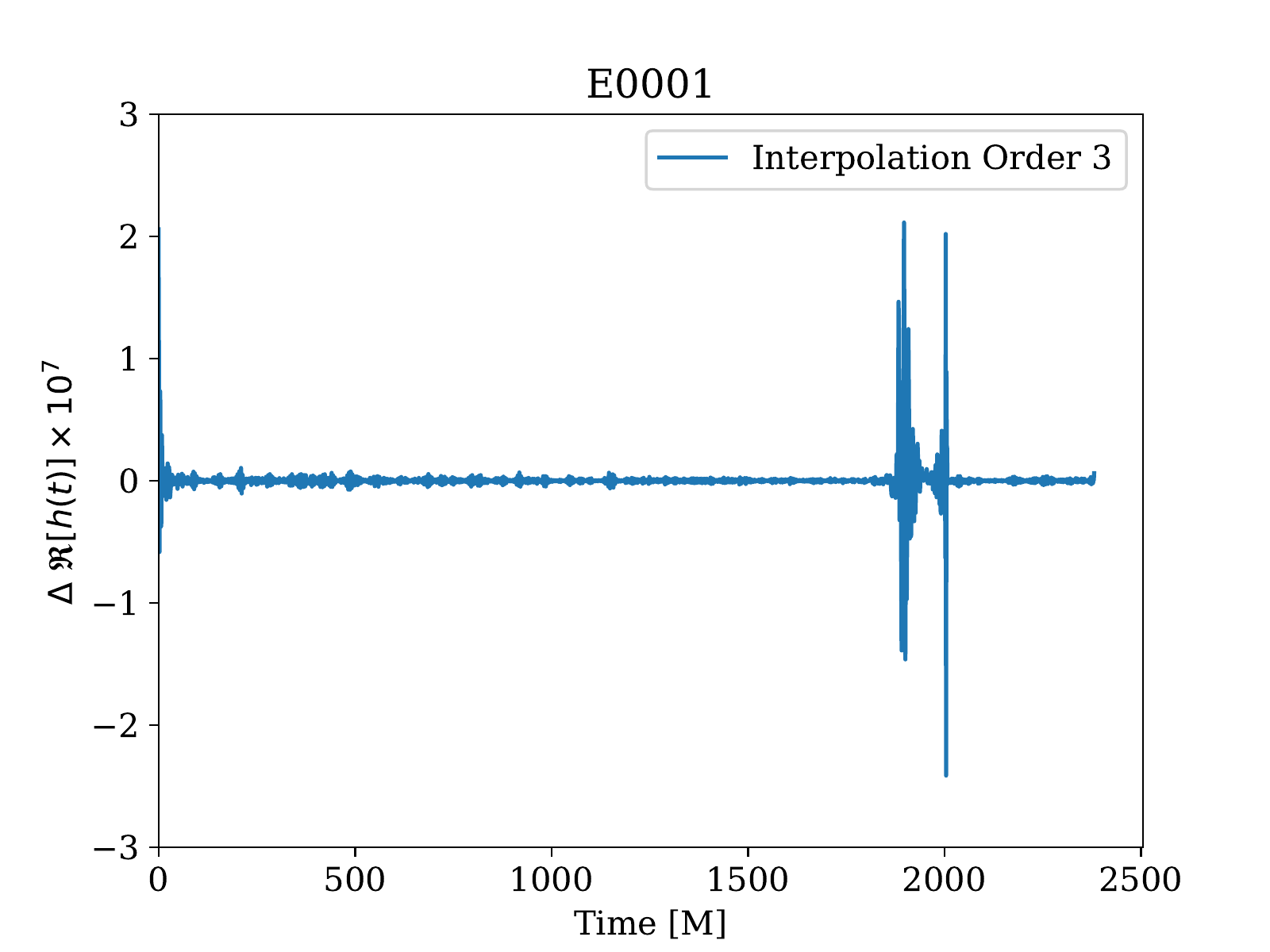}
				}
			\centerline{
				\includegraphics[height=0.4\textwidth,  clip]{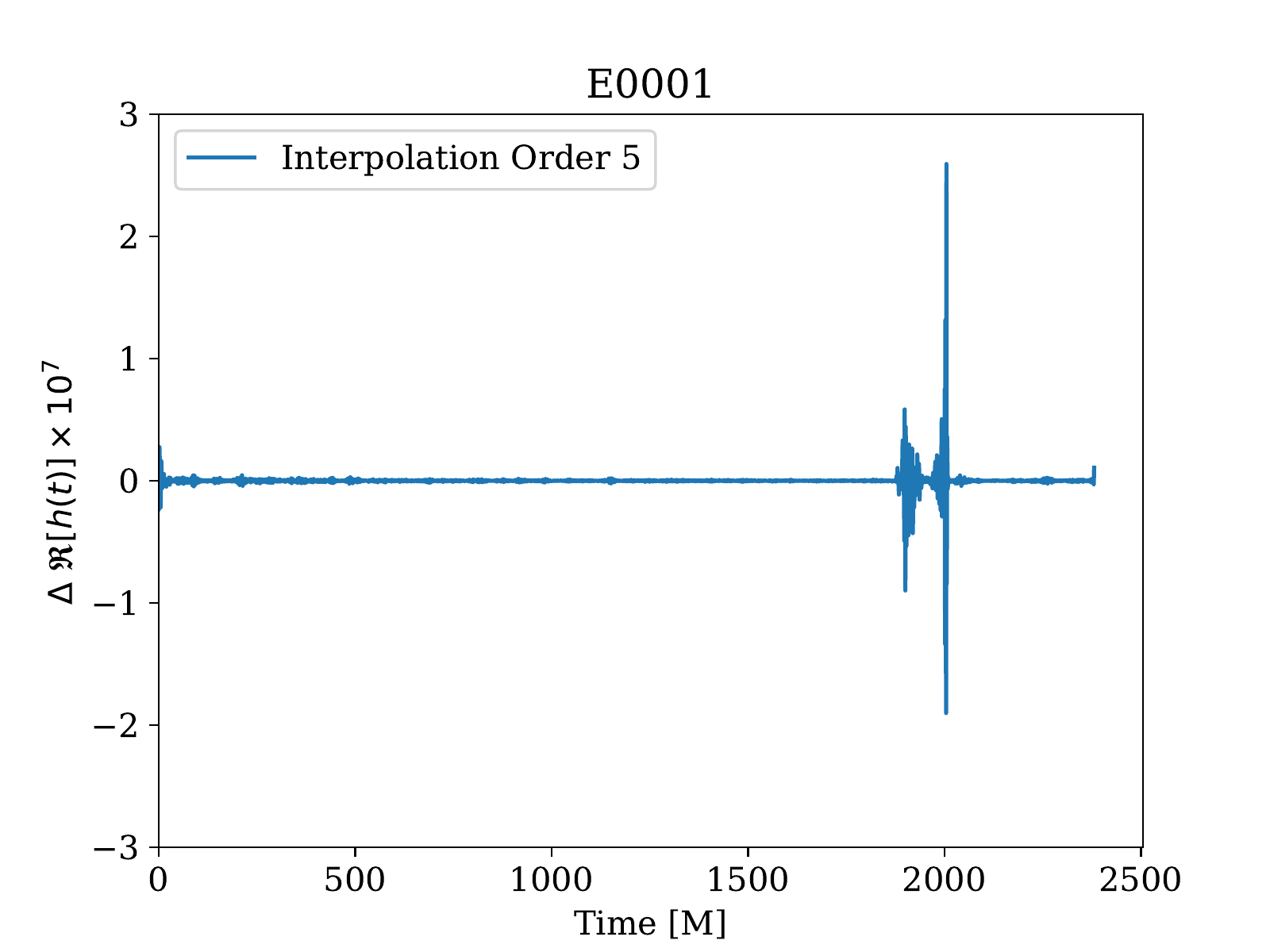}
				\includegraphics[height=0.4\textwidth,  clip]{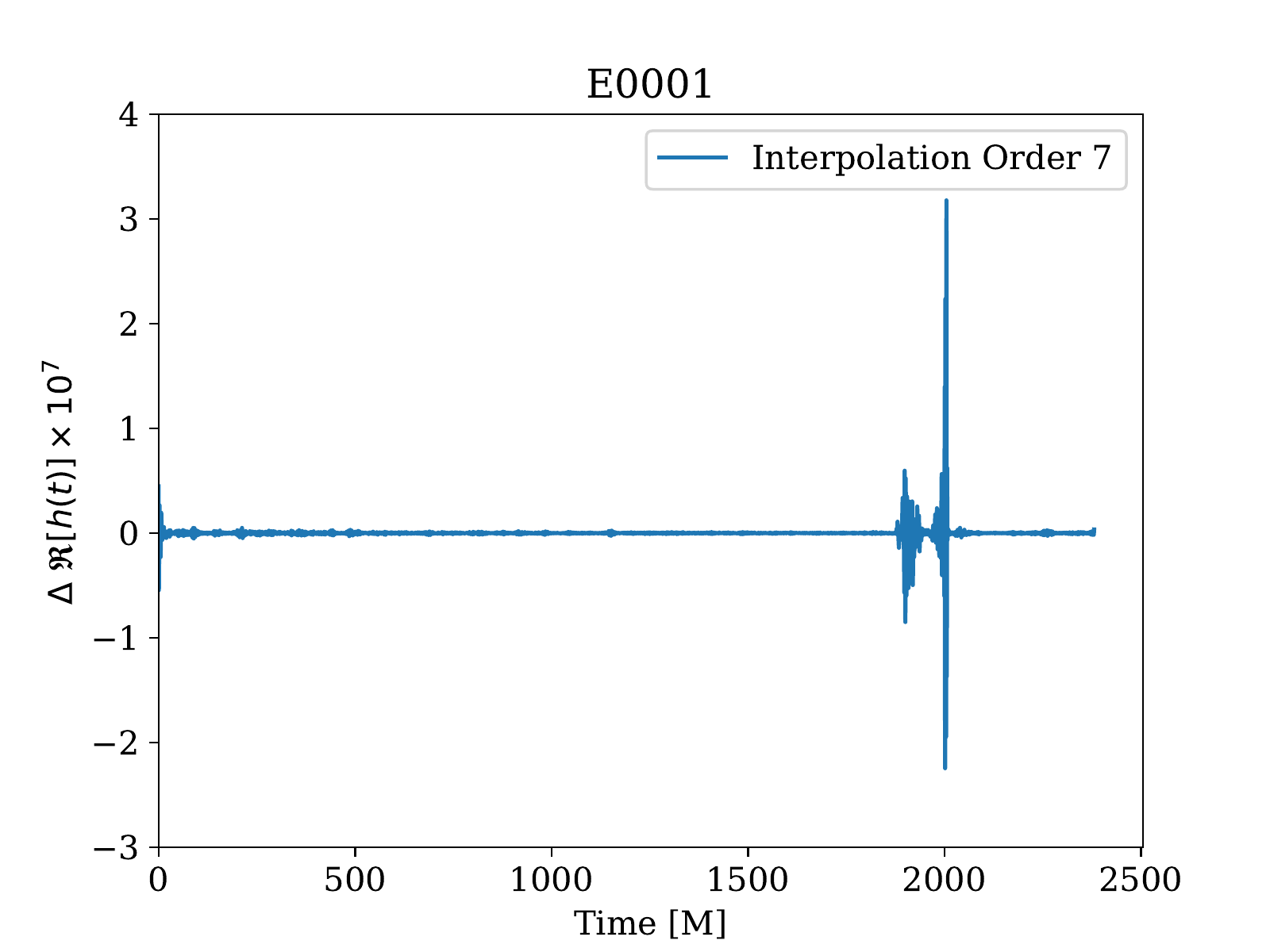}
			}	
				\caption{Discrepancy in the gravitational wave strain computed at future null infinity between \texttt{POWER} and \texttt{SimulationTools}: \( \Delta \Re\left[h(t)\right]\equiv \Re\left(h^{\texttt{POWER}} - h^{\texttt{SimTools}}\right)\). The interpolation orders used in \texttt{POWER} to produce \(h(t)\) are indicated in the panels. Note that the y-axes of the panels have been multiplied by \(10^4\) and \(10^{7}\) for interpolation orders $1$ and $\ge3$, respectively.}
				\label{fig_comparison_orders}
		\end{figure*}

\begin{figure*}[htp!]
		\centerline{
				\includegraphics[height=0.4\textwidth,  clip]{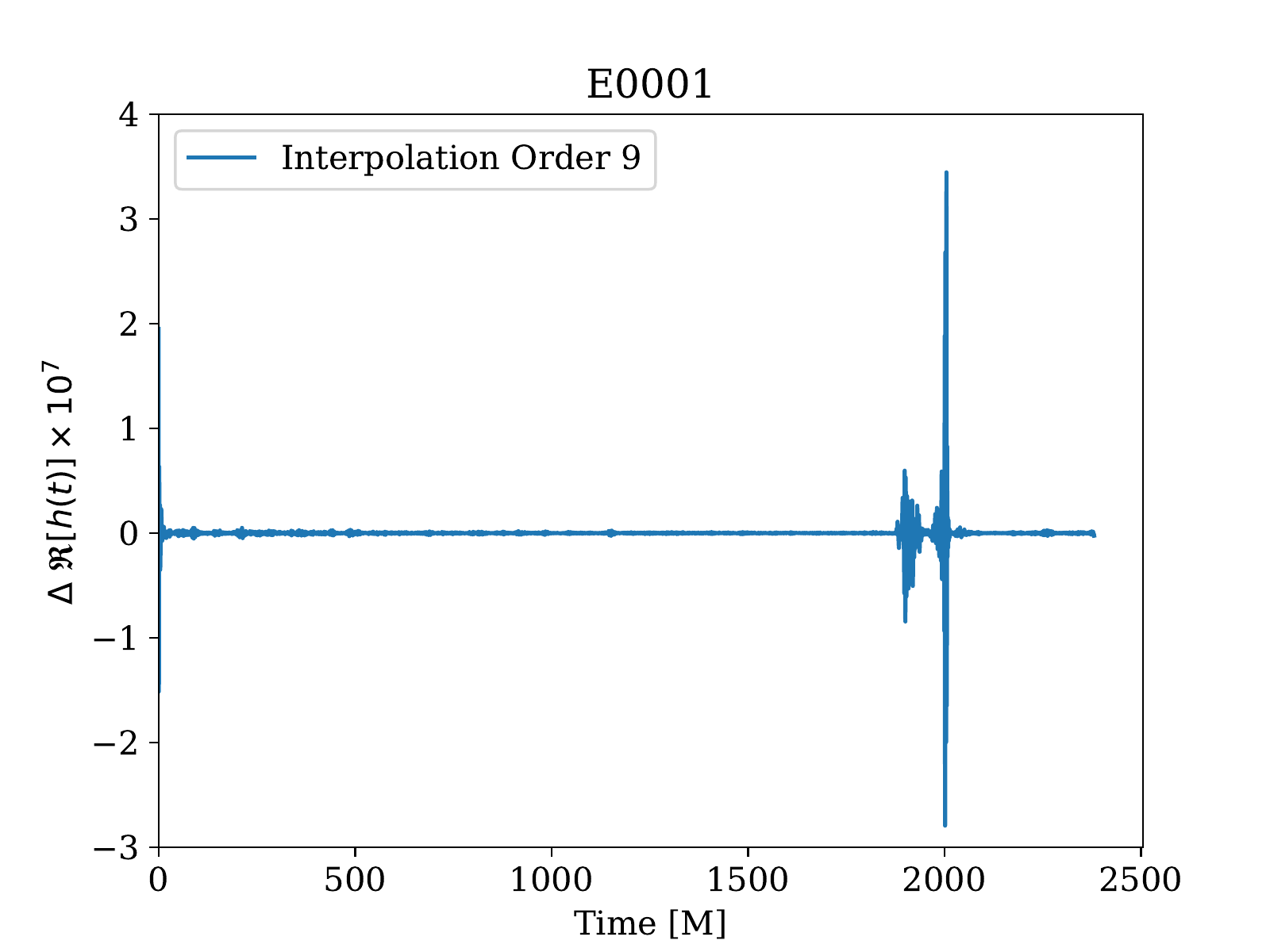}
				\includegraphics[height=0.4\textwidth,  clip]{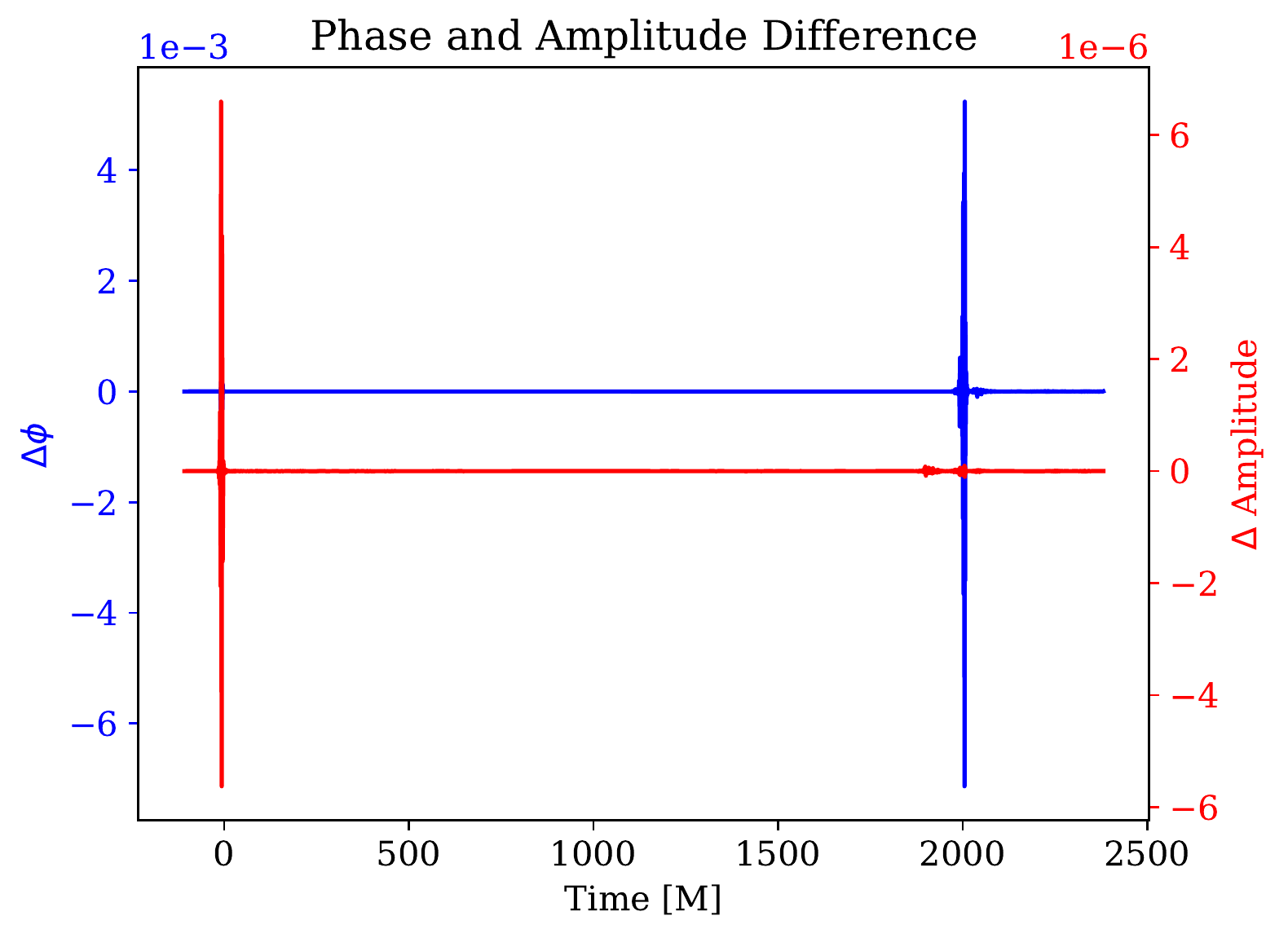}
				}
		\centerline{		
				\includegraphics[height=0.4\textwidth,  clip]{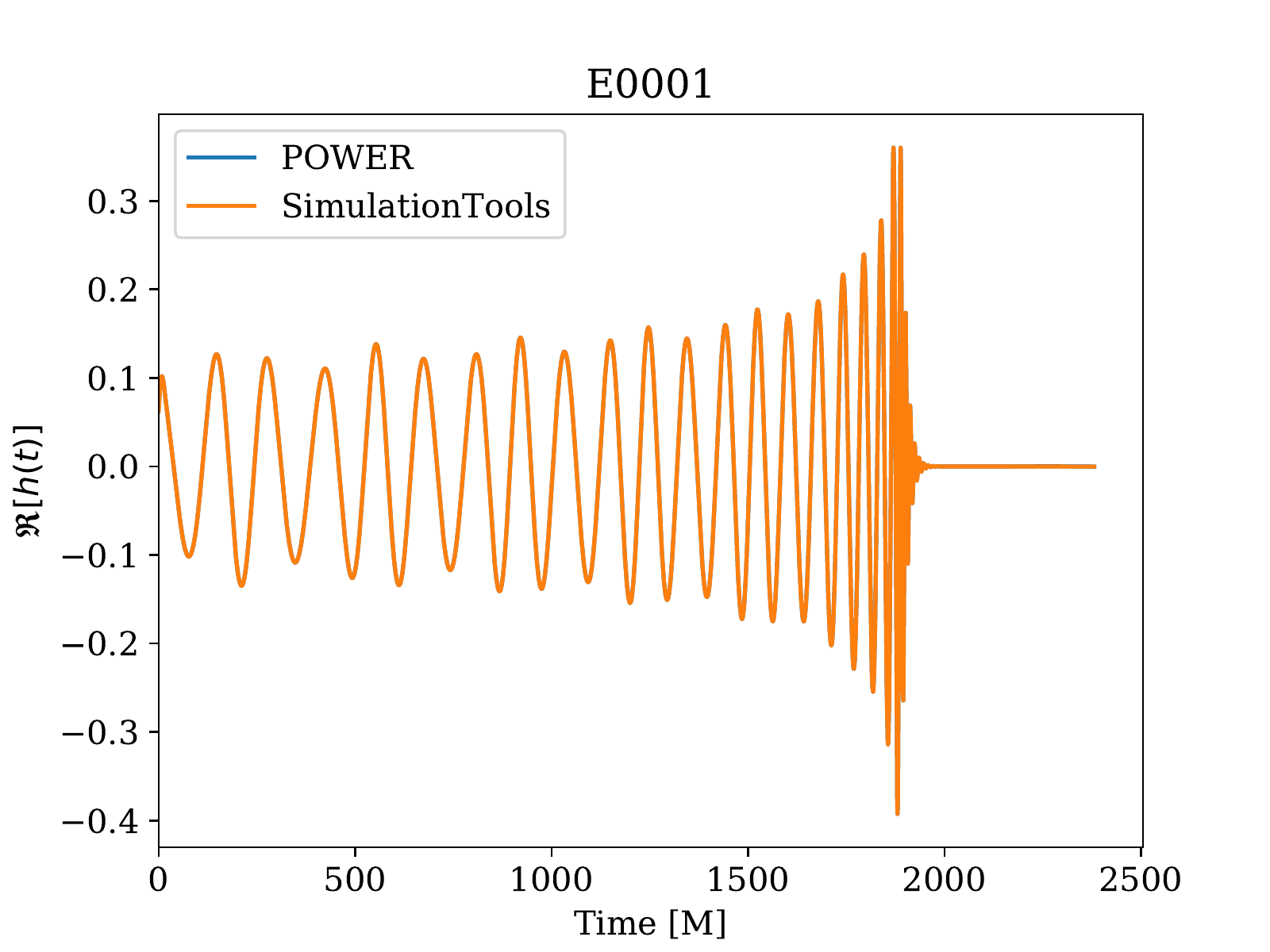}
			}
		\caption{Top left: As \Fref{fig_comparison_orders}, but now showing interpolation order 9. Top right: with interpolation order 9, we show the difference in phase and amplitude of the waveform strain extracted at future null infinity between \texttt{POWER} and \texttt{SimulationTools}. The bottom panel shows the real part of the waveform strain, \(\Re\left[h(t)\right]\), computed at future null infinity, between \texttt{POWER} and \texttt{SimulationTools} using interpolation order 9.}
				\label{fig_comparison_orders_two}
		\end{figure*}

\begin{figure*}[htp!]
		\centerline{
		\includegraphics[height=0.4\textwidth,  clip]{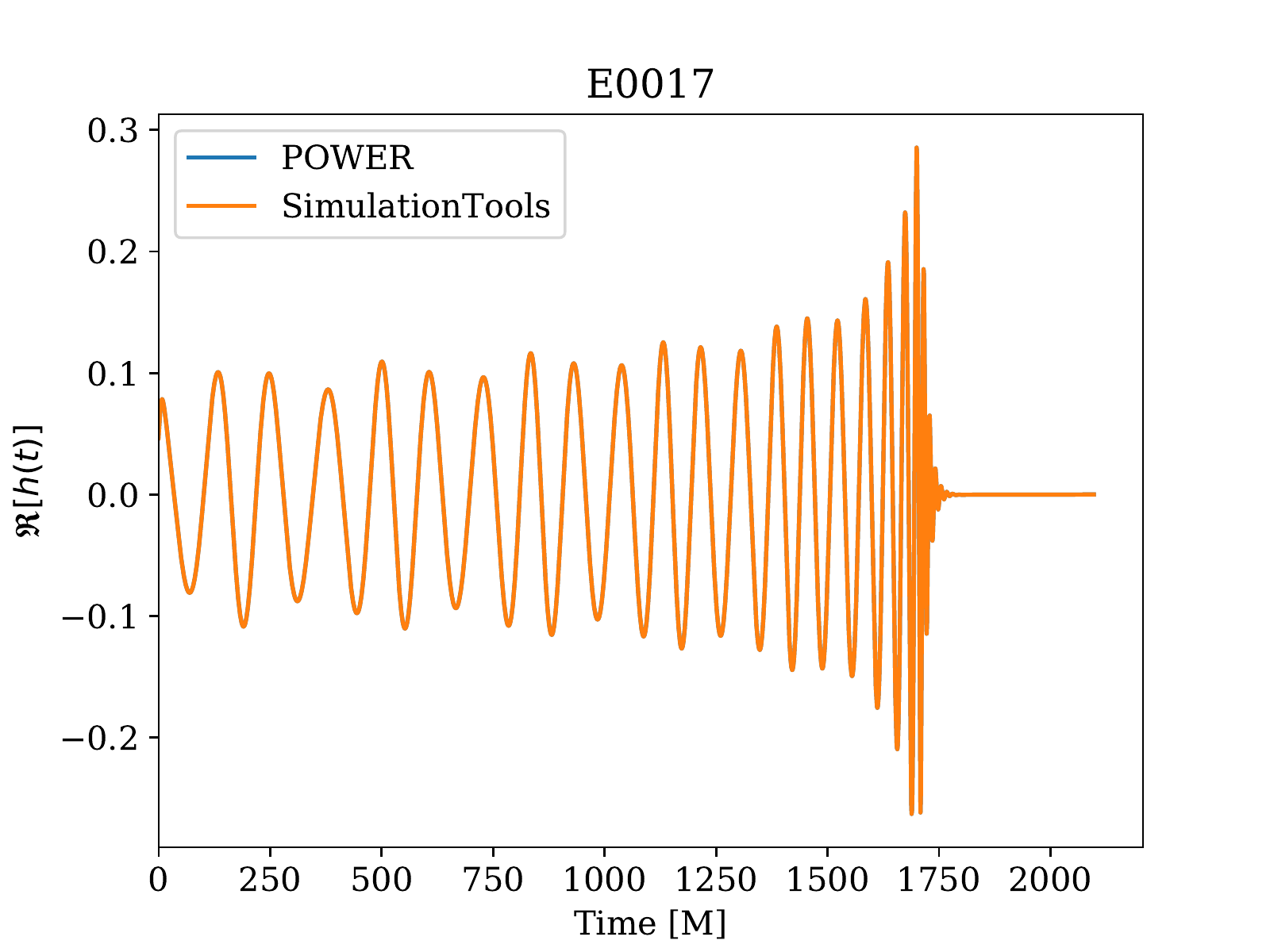}
		\includegraphics[height=0.4\textwidth,  clip]{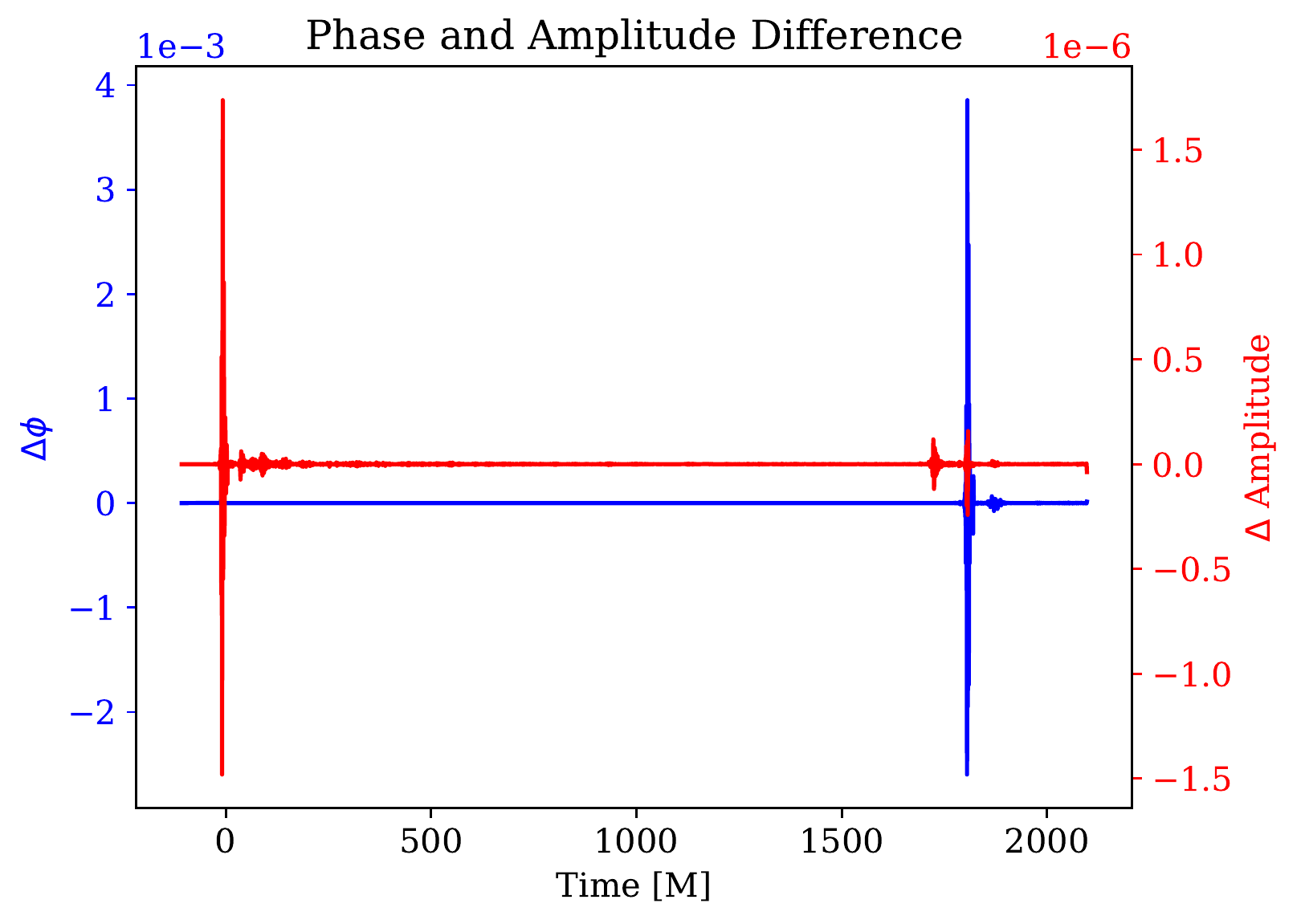}
		}
		\centerline{
		\includegraphics[height=0.4\textwidth,  clip]{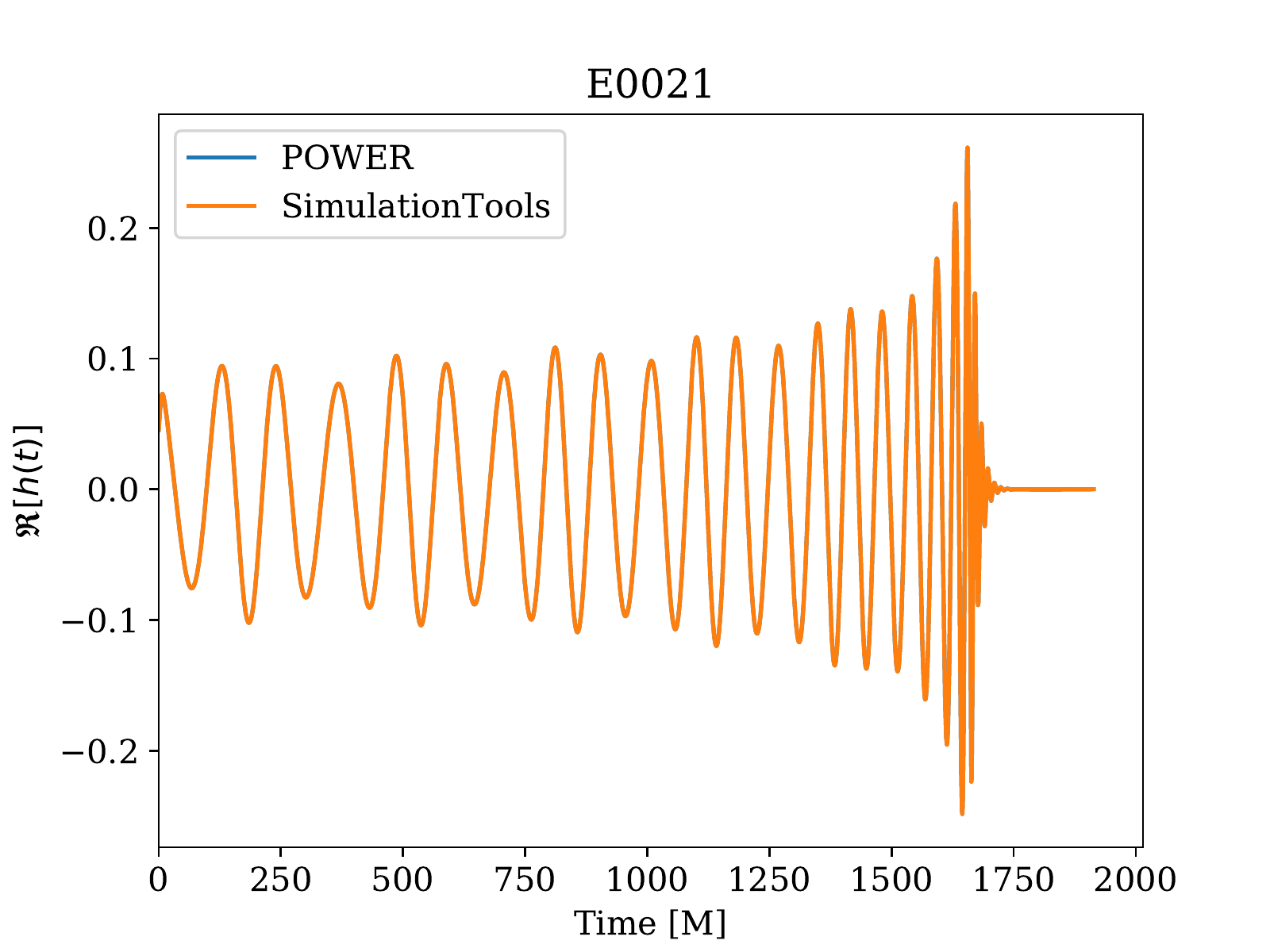}
		\includegraphics[height=0.4\textwidth,  clip]{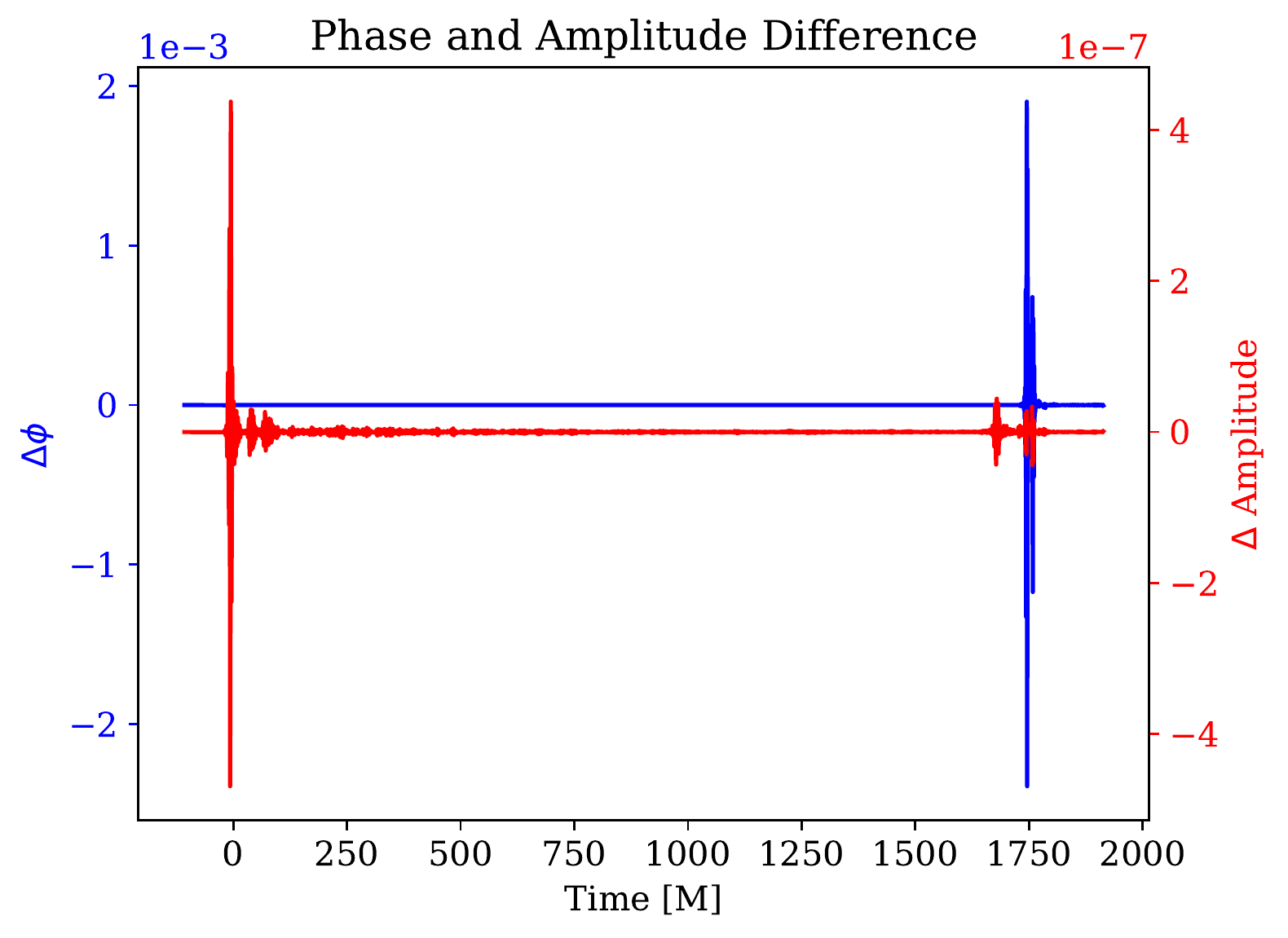}
		}
		\caption{Using interpolation order 9, we show the difference in phase and amplitude of the waveform strain extracted at future null infinity between \texttt{POWER} and \texttt{SimulationTools}. See \Tref{results} for details about the mass-ratio and eccentricities of these simulations.}
		\label{set_one}
	\end{figure*}

			 \noindent In \Fref{fig_comparison_orders}, \Fref{fig_comparison_orders_two} and \Fref{set_one}, we present a direct comparison of the gravitational wave strain computed at future null infinity by \texttt{POWER} and \texttt{SimulationTools}, \( \Delta \Re\left[h(t)\right]\equiv \Re\left(h^{\texttt{POWER}} - h^{\texttt{SimTools}}\right)\), for a set of moderately eccentric BBH systems. We also present the difference in amplitude and phase for the waveform strain extracted both from \texttt{POWER} and \texttt{SimulationTools}.

			 \texttt{SimulationTools} uses Hermite interpolation to produce \(h(t)\), whereas \texttt{POWER} uses \texttt{numpy}, which relies on spline interpolation. We have found that both methods agree, i.e, that \( \Delta \Re\left[h(t)\right]\leq 10^{-7}\), when we use interpolation order five or above. This result is robust for our catalog of more than \(70\) \texttt{Einstein Toolkit} BBH NR simulations~\cite{Huerta:ncsacatalog}. The user can modify this interpolation order to meet specific needs.
			 
			 It is worth emphasizing that in \texttt{POWER} the waveform phase and amplitude are interpolated to compute phase and amplitude at fixed retarded time $t_{\rm ret} = t-r$ for each extraction radii when radially extrapolating the strain. The interpolation is performed after the strain at the individual detectors is calculated, but before the radially extrapolated strain is computed. The interpolation order refers to spline interpolation order for \texttt{POWER} and Hermite interpolation order for \texttt{SimulationTools}. The examples shown only use the \((\ell,\,m)= (2,\,2)\) mode, but we show in Figure~\ref{orders} that \texttt{POWER} processes all of the modes present in the input files.
			 
			 The results shown in \Fref{fig_comparison_orders} and \Fref{set_one} are those of the highest resolution numerical simulation with the errors quoted being the difference between this value and the Richardson extrapolated value. These results are not Richardson extrapolated to infinite resolution directly though the numbers provided (highest level value plus error estimate) allow for a simple computation of the Richardson extrapolated value.

	 For the two NR simulations shown in \Fref{set_one}, we set the interpolation order to 9, and find that \( \Delta \Re\left[h(t)\right]\leq 10^{-7}\), and that discrepancies in phase and amplitude are always \(\leq (10^{-3}, \, 10^{-7})\), respectively, which is and less than the numerical error of the NR simulations.  It is reassuring that the differences \(( \Delta \Re\left[h(t)\right], \Delta \phi, \Delta {\rm Amplitude})\) for different mass-ratios and eccentricities are of similar magnitude for all waveforms in our catalog.

\section{Science cases and extension}
\label{ext}

To showcase the versatility and applicability of \texttt{POWER} both for vacuum and matter NR simulations, in this section we discuss a variety of projects that can make use of our new software package.

\begin{figure*}[htp!]
			\centerline{
				\includegraphics[height=0.7\textwidth,  clip]{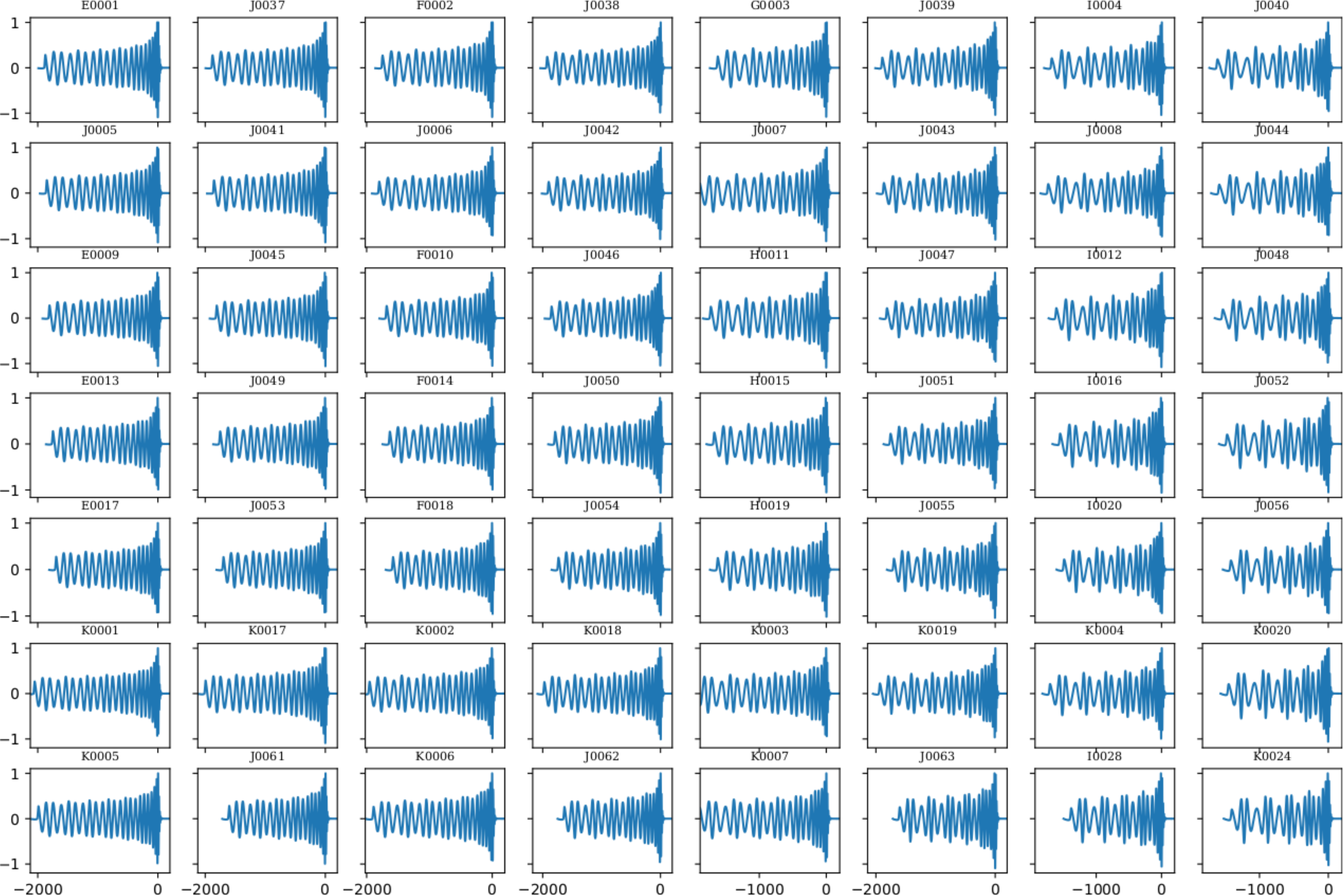}
				}
		
		\caption{Numerical relativity waveforms generated with our Blue Waters and XSEDE allocations, post-processed with our open source waveform extractor \texttt{POWER}. From top to bottom, each row \(i\) presents BBH mergers of mass-ratio \(q=1+0.5\times i\) with \(i=\{0,\ldots,6\}\). The properties of these simulations are discussed in detailed in an accompanying paper~\cite{Huerta:ncsacatalog}.}
			\label{our_catalog}
		\end{figure*}

We have recently used \texttt{POWER} to post-process our first catalog of eccentric NR simulations that span a parameter space that describes BBH mergers with mass-ratios \(q\leq 6\), and eccentricities \(e\leq0.21\) ten orbits before merger. The first batch of waveforms with mass-ratios \(q\leq4\) in shown in \Fref{our_catalog}. A detailed analysis of these NR simulations is presented in an accompanying paper~\cite{Huerta:ncsacatalog}.

\begin{figure*}[htp!]
			\centerline{
				\includegraphics[height=0.4\textwidth,  clip]{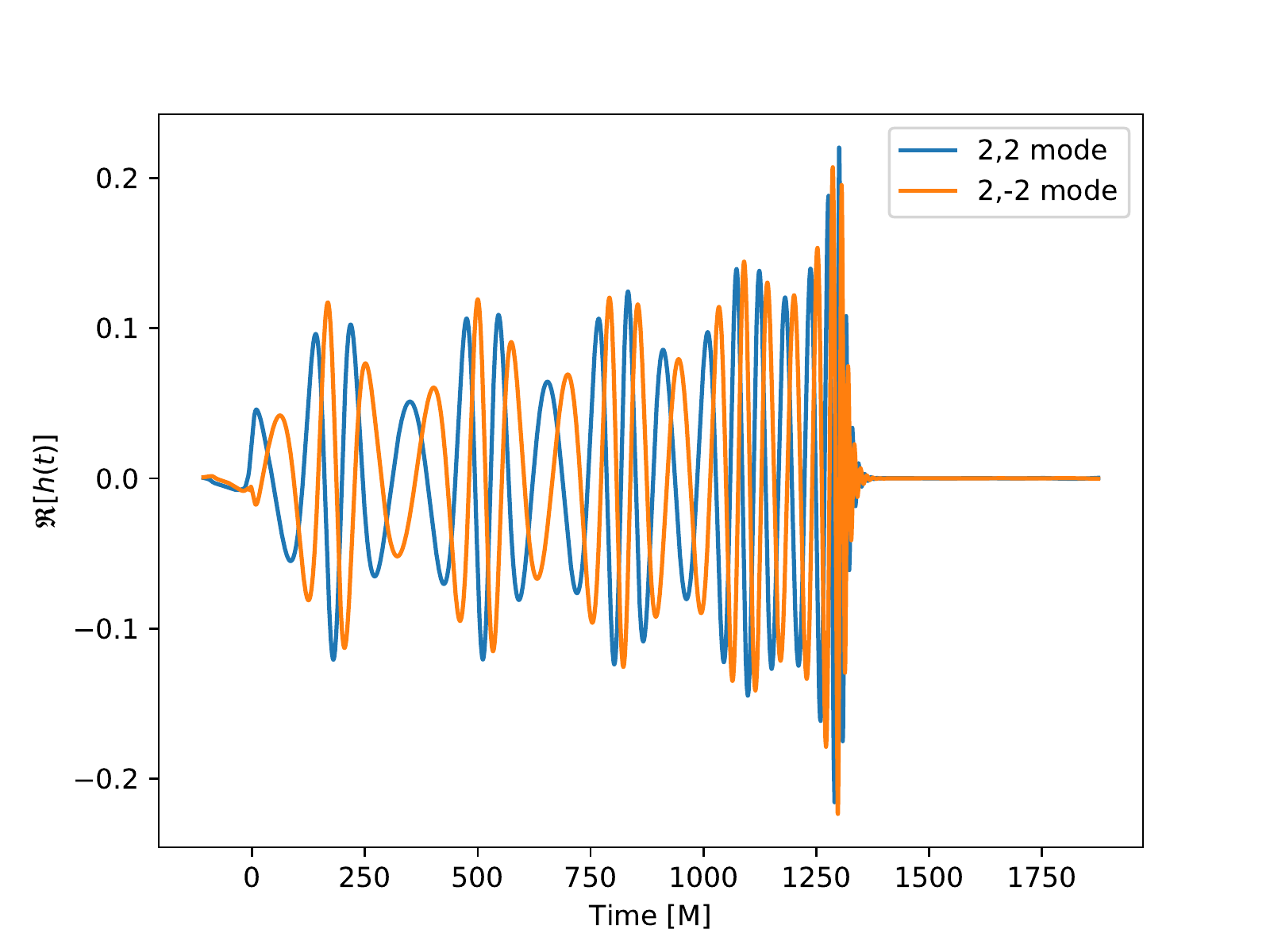}
				\includegraphics[height=0.4\textwidth,  clip]{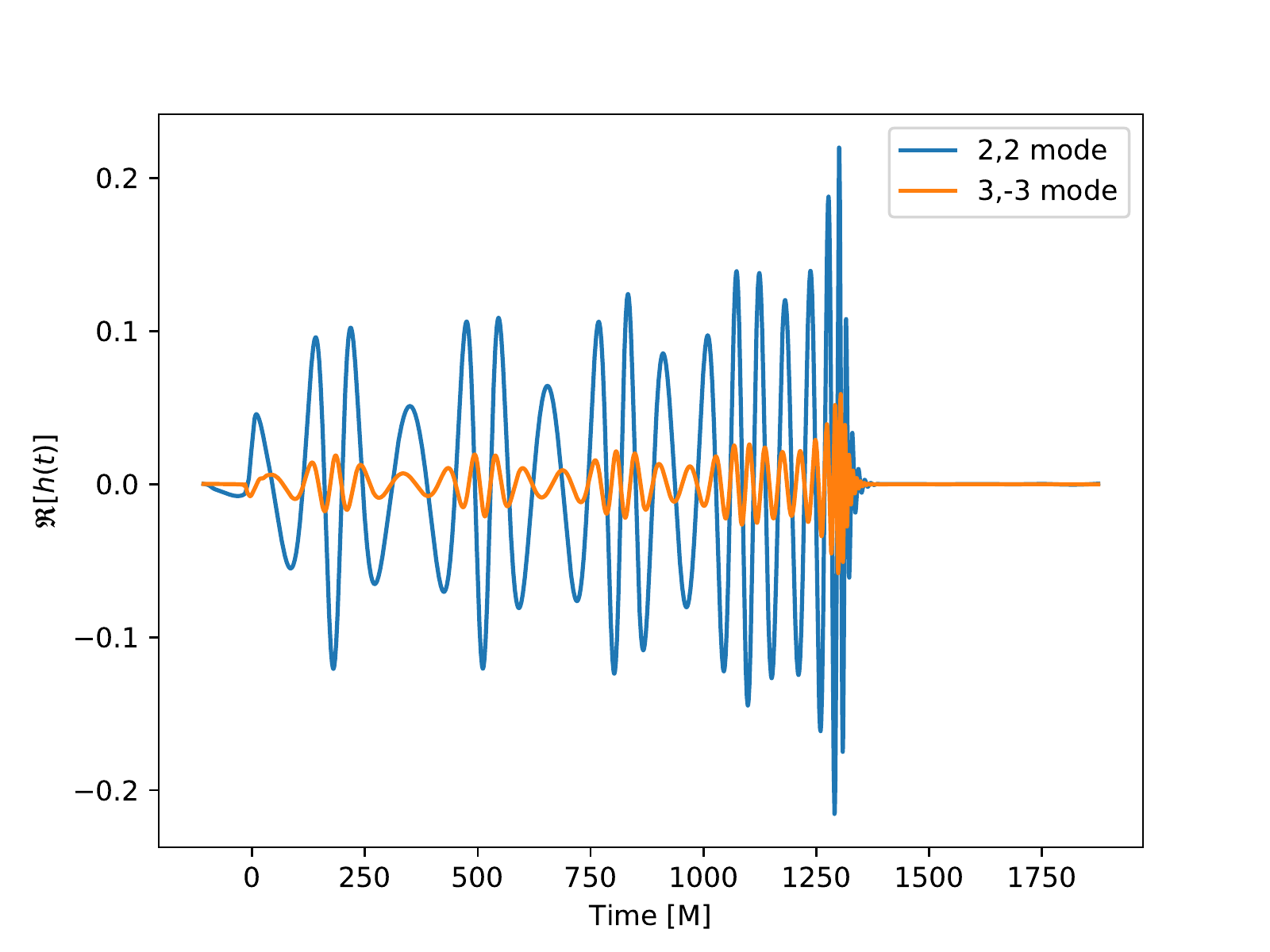}
			}
			\centerline{
				\includegraphics[height=0.4\textwidth,  clip]{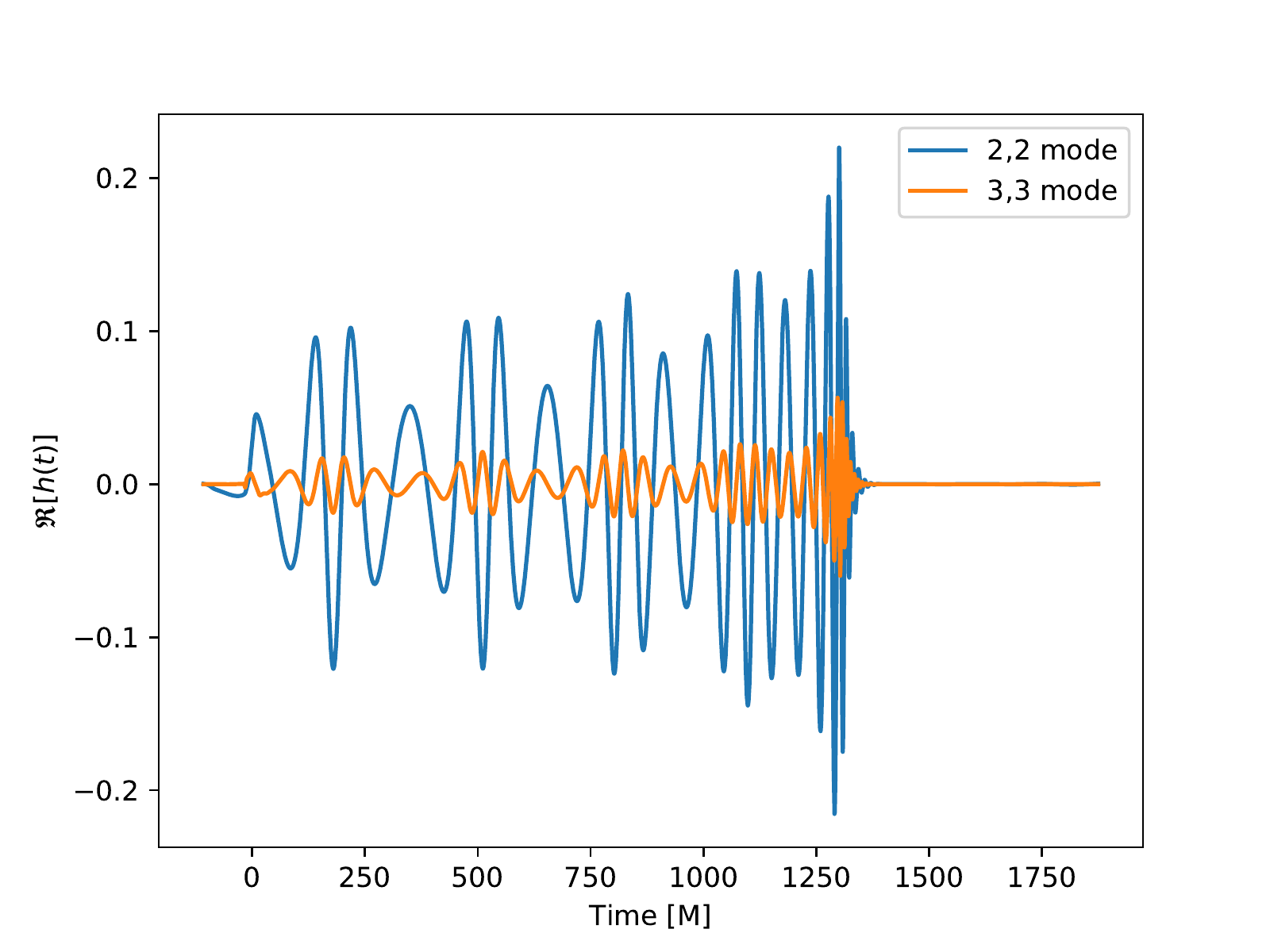}
				\includegraphics[height=0.4\textwidth,  clip]{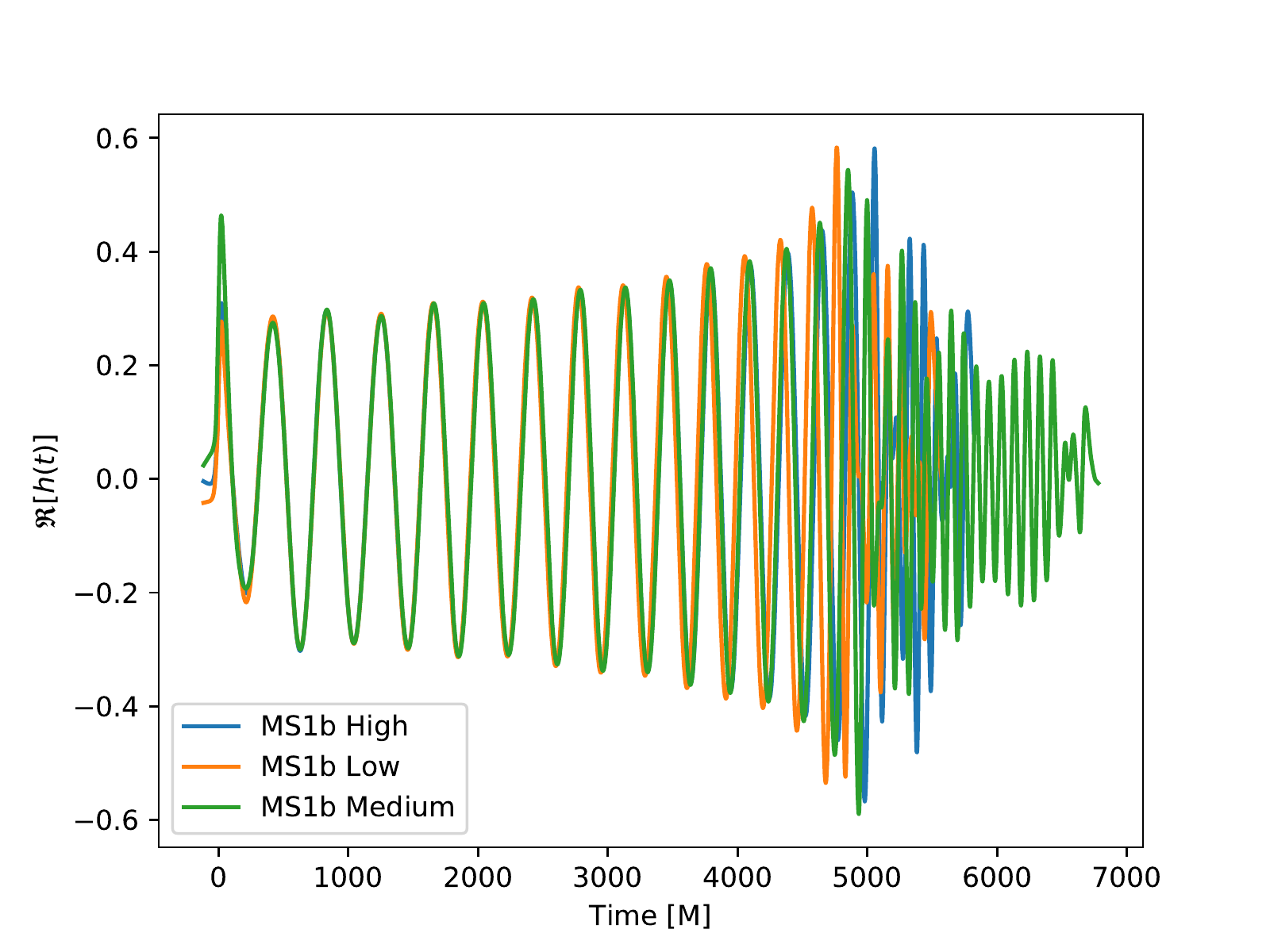}
			}
			\caption{Top panels, and bottom left panel: extraction of higher-order waveform multipoles from the binary black hole NR simulation J0068. Bottom right panel: extraction of the waveform strain of a binary neutron star merger simulated with the \texttt{GRHydro} code, and using the MS1b equation of state.}
			\label{orders}
		\end{figure*}

\noindent Given that \texttt{POWER} is capable of extracting all of the higher-order waveform multipoles present in the data products of NR simulations, we have exploited this feature to start extracting the higher-order modes of our eccentric BBH merger catalog. These results are shown in \Fref{orders}. These results can shed light into the importance of these higher-order contributions for the detectability of eccentric compact binary coalescence, and issue that has not been addressed in the literature.

Finally, even though \texttt{POWER} was developed to post-process vacuum NR simulations, we have now made it flexible enough to post-process the data products of matter simulations. In the bottom right panel of \Fref{orders}, we present the waveform strain computed at future null infinity from the merger of two equal mass neutron stars, using the MS1b equation of state. This simulation was generated and post-processed using the open source \texttt{GRHydro} code~\cite{2014CQGra..31a5005M}, and \texttt{POWER}, respectively, in the Blue Waters supercomputer.

		In summary, we have developed an open source tool that is ready to use to monitor the progress of NR campaigns, and to compute the GW strain \(h(t)\) at future null infinity. These results indicate that \texttt{POWER} is an ideal open source tool to post-process NR simulations on any leadership HPC facility that provides \texttt{numpy} and \texttt{scipy}.
		
	\section{Conclusion}
	\label{end}	
	
	We have developed, and made publicly available, a \texttt{Python} package to monitor and post-process NR simulations on HPC environments. This software is available as a stand alone tool, and will be included in upcoming  releases of the open source, community software the \texttt{Einstein Toolkit}. This tool will facilitate and streamline the continuous monitoring needed in NR campaigns, and the subsequent post-processing required to compute the GW strain at future null infinity. This work fills in a critical void in the arsenal of tools that is available to numerical relativists interested in assisting the LIGO Scientific Collaboration in the validation of GW transients, or to develop new waveform models from the ground up, among other activities.
				
	\section{Acknowledgements}
	\label{ack}
	This research is part of the Blue Waters sustained-petascale computing project, which is supported by the National Science Foundation (awards OCI-0725070 and ACI-1238993) and the State of Illinois. Blue Waters is a joint effort of the University of Illinois at Urbana-Champaign and its National Center for Supercomputing Applications (NCSA). The eccentric numerical relativity simulations used in this article were generated with the open source, community software, the Einstein Toolkit on the Blue Waters petascale supercomputer and XSEDE (TG-PHY160053).  We acknowledge support from the NCSA and the SPIN (Students Pushing Innovation) Program at NCSA. We thank Ian Hinder and Barry Wardell for the \texttt{SimulationTools} analysis package. Plots were generated with \texttt{Matplotlib}~\cite{Hunter:2007}. \texttt{POWER} uses \texttt{numpy}~\cite{Walt:2011} and \texttt{scipy}~\cite{Jones:2001}.

	\vspace{2pc}
	
	\bibliographystyle{iopart-num}
	\bibliography{references}

\providecommand{\newblock}{}
\begin{thebibliography}{10}
\expandafter\ifx\csname url\endcsname\relax
  \def\url#1{{\tt #1}}\fi
\expandafter\ifx\csname urlprefix\endcsname\relax\def\urlprefix{URL }\fi
\providecommand{\eprint}[2][]{\url{#2}}

\bibitem{gr}
{Einstein} A 1915 {\em K{\"o}niglich Preussische Akademie der Wissenschaften Zu
  Berlin, Sitzungberichte\/} {\bf 1915} 844--847

\bibitem{preto}
{Pretorius} F 2005 {\em Physical Review Letters\/} {\bf 95} 121101--+
  (\textit{Preprint} \eprint{arXiv:gr-qc/0507014})

\bibitem{naka:1987}
{Nakamura} T, {Oohara} K and {Kojima} Y 1987 {\em Progress of Theoretical
  Physics Supplement\/} {\bf 90} 1--218

\bibitem{shiba:1995}
{Shibata} M and {Nakamura} T 1995 {\em \prd\/} {\bf 52} 5428--5444

\bibitem{baum:1999}
{Baumgarte} T~W and {Shapiro} S~L 1999 {\em \prd\/} {\bf 59} 024007
  (\textit{Preprint} \eprint{gr-qc/9810065})

\bibitem{baker:2006}
{Baker} J~G, {Centrella} J, {Choi} D~I, {Koppitz} M and {van Meter} J 2006 {\em
  Physical Review Letters\/} {\bf 96} 111102 (\textit{Preprint}
  \eprint{gr-qc/0511103})

\bibitem{camp:2006}
{Campanelli} M, {Lousto} C~O, {Marronetti} P and {Zlochower} Y 2006 {\em
  Physical Review Letters\/} {\bf 96} 111101 (\textit{Preprint}
  \eprint{gr-qc/0511048})

\bibitem{SathyaLRR:2009}
{Sathyaprakash} B~S and {Schutz} B~F 2009 {\em Living Reviews in Relativity\/}
  {\bf 12} 2 (\textit{Preprint} \eprint{0903.0338})

\bibitem{DI:2016}
{Abbott} B~P, {Abbott} R, {Abbott} T~D, {Abernathy} M~R, {Acernese} F, {Ackley}
  K, {Adams} C, {Adams} T, {Addesso} P, {Adhikari} R~X and et~al 2016 {\em
  Physical Review Letters\/} {\bf 116} 061102 (\textit{Preprint}
  \eprint{1602.03837})

\bibitem{secondBBH:2016}
{Abbott} B~P, {Abbott} R, {Abbott} T~D, {Abernathy} M~R, {Acernese} F, {Ackley}
  K, {Adams} C, {Adams} T, {Addesso} P, {Adhikari} R~X and et~al 2016 {\em
  Physical Review Letters\/} {\bf 116} 241103 (\textit{Preprint}
  \eprint{1606.04855})

\bibitem{thirddetection}
{Abbott} B~P, {Abbott} R, {Abbott} T~D, {Abernathy} M~R, {Acernese} F, {Ackley}
  K, {Adams} C, {Adams} T, {Addesso} P, {Adhikari} R~X and et~al 2017 {\em
  Physical Review Letters\/} {\bf 118}(22) 221101
  \urlprefix\url{https://link.aps.org/doi/10.1103/PhysRevLett.118.221101}

\bibitem{2017arXiv170909660T}
{The LIGO Scientific Collaboration}, {the Virgo Collaboration}, {Abbott} B~P,
  {Abbott} R, {Abbott} T~D, {Acernese} F, {Ackley} K, {Adams} C, {Adams} T,
  {Addesso} P and et~al 2017 {\em ArXiv e-prints\/} ArXiv:1709.09660 [gr-qc]
  (\textit{Preprint} \eprint{1709.09660})

\bibitem{Tara:2014}
{Taracchini} A, {Buonanno} A, {Pan} Y, {Hinderer} T, {Boyle} M, {Hemberger}
  D~A, {Kidder} L~E, {Lovelace} G, {Mrou{\'e}} A~H, {Pfeiffer} H~P, {Scheel}
  M~A, {Szil{\'a}gyi} B, {Taylor} N~W and {Zenginoglu} A 2014 {\em \prd\/} {\bf
  89} 061502 (\textit{Preprint} \eprint{1311.2544})

\bibitem{khan:2016PhRvD}
{Khan} S, {Husa} S, {Hannam} M, {Ohme} F, {P{\"u}rrer} M, {Forteza} X~J and
  {Boh{\'e}} A 2016 {\em \prd\/} {\bf 93} 044007 (\textit{Preprint}
  \eprint{1508.07253})

\bibitem{2016arXiv161107531T}
{The LIGO Scientific Collaboration}, {the Virgo Collaboration}, {Abbott} B~P,
  {Abbott} R, {Abbott} T~D, {Abernathy} M~R, {Acernese} F, {Ackley} K, {Adams}
  C, {Adams} T and et~al 2016 {\em ArXiv e-prints\/} (\textit{Preprint}
  \eprint{1611.07531})

\bibitem{NRI:2016}
{Abbott} B~P, {Abbott} R, {Abbott} T~D, {Abernathy} M~R, {Acernese} F, {Ackley}
  K, {Adams} C, {Adams} T, {Addesso} P, {Adhikari} R~X and et~al 2016 {\em
  \prd\/} {\bf 94} 064035 (\textit{Preprint} \eprint{1606.01262})

\bibitem{Lehner:2014a}
{Lehner} L and {Pretorius} F 2014 {\em \araa\/} {\bf 52} 661--694
  (\textit{Preprint} \eprint{1405.4840})

\bibitem{scenarioligo:2016LRR}
{Abbott} B~P, {Abbott} R, {Abbott} T~D, {Abernathy} M~R, {Acernese} F, {Ackley}
  K, {Adams} C, {Adams} T, {Addesso} P, {Adhikari} R~X and et~al 2016 {\em
  Living Reviews in Relativity\/} {\bf 19} (\textit{Preprint}
  \eprint{1304.0670})

\bibitem{2017JCoPh.335...84K}
{Kidder} L~E, {Field} S~E, {Foucart} F, {Schnetter} E, {Teukolsky} S~A, {Bohn}
  A, {Deppe} N, {Diener} P, {H{\'e}bert} F, {Lippuner} J, {Miller} J, {Ott}
  C~D, {Scheel} M~A and {Vincent} T 2017 {\em Journal of Computational
  Physics\/} {\bf 335} 84--114 (\textit{Preprint} \eprint{1609.00098})

\bibitem{2016PhRvD..93l4062H}
{Haas} R, {Ott} C~D, {Szilagyi} B, {Kaplan} J~D, {Lippuner} J, {Scheel} M~A,
  {Barkett} K, {Muhlberger} C~D, {Dietrich} T, {Duez} M~D, {Foucart} F,
  {Pfeiffer} H~P, {Kidder} L~E and {Teukolsky} S~A 2016 {\em \prd\/} {\bf 93}
  124062 (\textit{Preprint} \eprint{1604.00782})

\bibitem{2015CQGra..32q5009E}
{Etienne} Z~B, {Paschalidis} V, {Haas} R, {M{\"o}sta} P and {Shapiro} S~L 2015
  {\em Classical and Quantum Gravity\/} {\bf 32} 175009 (\textit{Preprint}
  \eprint{1501.07276})

\bibitem{2014CQGra..31a5005M}
{M{\"o}sta} P, {Mundim} B~C, {Faber} J~A, {Haas} R, {Noble} S~C, {Bode} T,
  {L{\"o}ffler} F, {Ott} C~D, {Reisswig} C and {Schnetter} E 2014 {\em
  Classical and Quantum Gravity\/} {\bf 31} 015005 (\textit{Preprint}
  \eprint{1304.5544})

\bibitem{ETL:2012CQGra}
{L{\"o}ffler} F, {Faber} J, {Bentivegna} E, {Bode} T, {Diener} P, {Haas} R,
  {Hinder} I, {Mundim} B~C, {Ott} C~D, {Schnetter} E, {Allen} G, {Campanelli} M
  and {Laguna} P 2012 {\em Classical and Quantum Gravity\/} {\bf 29} 115001
  (\textit{Preprint} \eprint{1111.3344})

\bibitem{simulationtools-web}
Hinder I and Wardell B Simulationtools is a free software package for the
  analysis of numerical simulation data in mathematica.
  http://simulationtools.org/

\bibitem{etweb}
{The Einstein Toolkit} \url{http://einsteintoolkit.org}

\bibitem{hdf5-web}
{The HDF Group} 1997-2017 {Hierarchical Data Format, version 5}
  http://www.hdfgroup.org/HDF5/

\bibitem{reis:2011CQG}
{Reisswig} C and {Pollney} D 2011 {\em Classical and Quantum Gravity\/} {\bf
  28} 195015 (\textit{Preprint} \eprint{1006.1632})

\bibitem{Huerta:ncsacatalog}
{Huerta} E~A {\em et~al.\/} 2017

\bibitem{Hunter:2007}
Hunter J~D 2007 {\em Computing in Science Engineering\/} {\bf 9} 90--95 ISSN
  1521-9615

\bibitem{Walt:2011}
van~der Walt S, Colbert S~C and Varoquaux G 2011 {\em Computing in Science
  Engineering\/} {\bf 13} 22--30 ISSN 1521-9615

\bibitem{Jones:2001}
Jones E, Oliphant T, Peterson P {\em et~al.\/} 2001--2017 {SciPy}: Open source
  scientific tools for {Python} [Online; accessed \today]
  \urlprefix\url{http://www.scipy.org/}

\end{thebibliography}

\end{document}